%% file: main.tex
\documentclass[aps,prl,twocolumn,superscriptaddress,10pt]{revtex4-2}
\usepackage{kotex}
\usepackage{graphicx}
\usepackage{color}
\usepackage[normalem]{ulem}
\usepackage{amsmath}
\usepackage{mathtools}
\usepackage{natbib}
\usepackage{hyperref}
\usepackage{float}
\usepackage{multirow}
\usepackage{array}
\usepackage{xcolor}	
\usepackage{color, colortbl}
\usepackage{booktabs}
\hypersetup{
    colorlinks=true,
    linkcolor=blue,
    citecolor = blue,
    filecolor=magenta,      
    urlcolor=blue,
    pdfpagemode=FullScreen,
    }

\begin{document}

\title{The power of entanglement in distributed quantum machine learning}

\author{Yerim Kim}
\affiliation{Department of Physics, Korea University, Seoul 02841, Republic of Korea}
\author{Kiwmann Hwang}
\affiliation{Department of Physics, Korea University, Seoul 02841, Republic of Korea}
\author{Hyukjoon Kwon}
\affiliation{School of Computational Sciences, Korea Institute for Advanced Study, Seoul 02455, Republic of Korea}
\author{Yosep Kim}
\email{yosep9201@gmail.com}
\affiliation{Department of Physics, Korea University, Seoul 02841, Republic of Korea}
\begin{abstract}

The quantum internet aims to interconnect distant devices and enable large-scale computation through distributed quantum algorithms. One of the key obstacles is communication latency during computation. Even separations of a few hundred kilometers introduce millisecond-scale delays, which exceed the coherence times of many solid-state qubit platforms. In contrast, entanglement can be established beforehand and used as a practical resource to reduce communication complexity between remote nodes. Here we examine the utility of entanglement in distributed quantum machine learning for binary classification tasks. Drawing an analogy with the CHSH game, we show that entanglement improves classification accuracy across all datasets considered. We also find that excessive entanglement may degrade performance by reducing the effective dimension of the parameter space. This highlights the importance of using an appropriate amount and structure of entanglement in data embedding. Our findings bridge nonlocality and machine-learning advantage, providing a pathway toward distributed quantum computation beyond coherence-time constraints.

\end{abstract}
\maketitle

\begin{figure}[t!]
\centering
\includegraphics[width=0.95\columnwidth]{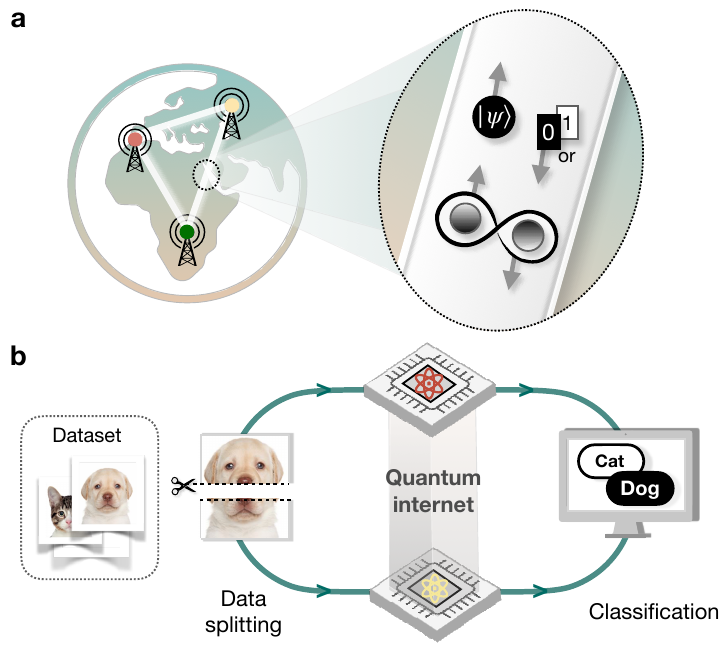}

\caption{\textbf{Conceptual framework.} 
\textbf{a}, Quantum internet based on communication resources, including qubits, classical bits, and shared entanglement (e.g., Bell pairs).
\textbf{b}, Distributed quantum machine learning for data classification. Input data exceeding the capacity of a single processor are partitioned and embedded into separate processors, and the resulting outputs are used for training or classification.}
\label{fig1}
\end{figure}

\noindent{\fontsize{11.2}{13}\selectfont\textbf{Introduction}}\vspace{0.3em}\\ 
\noindent Large-scale quantum computing remains constrained by the limited number of qubits and short coherence times, hindering the realization of quantum advantage~\cite{preskill2018NISQ,bharti2022noisy}. Distributed quantum computing offers a route to scalability by leveraging quantum internet and multiple cooperating processors~\cite{caleffi2024distributed,barral2025review,kimble2008quantuminternet, wehner2018quantuminternet,lee2026distributed, hwang2025DQML}. Quantum processors have been experimentally linked across various platforms over distances ranging from laboratory to metropolitan scales~\cite{aghaee2025scaling, main2025distributed,carrera2024combining,stolk2024metropolitan,knaut2024entanglement,zhou2025kilometer,yam2025cryogenicMicrowaveLink}. However, extending these networks to global scales would introduce millisecond-to-hundreds-of-milliseconds communication delays in both quantum and classical channels (Fig.~\ref{fig1}\textbf{a}). As the delays may exceed qubit coherence times, protocols rely on direct quantum state transfer or classical feedforward become infeasible, including quantum state and gate teleportation~\cite{bennett1993teleporting,gottesman1999demonstrating}.

\begin{figure*}[t!]
\centering
\includegraphics[width=0.77\linewidth]{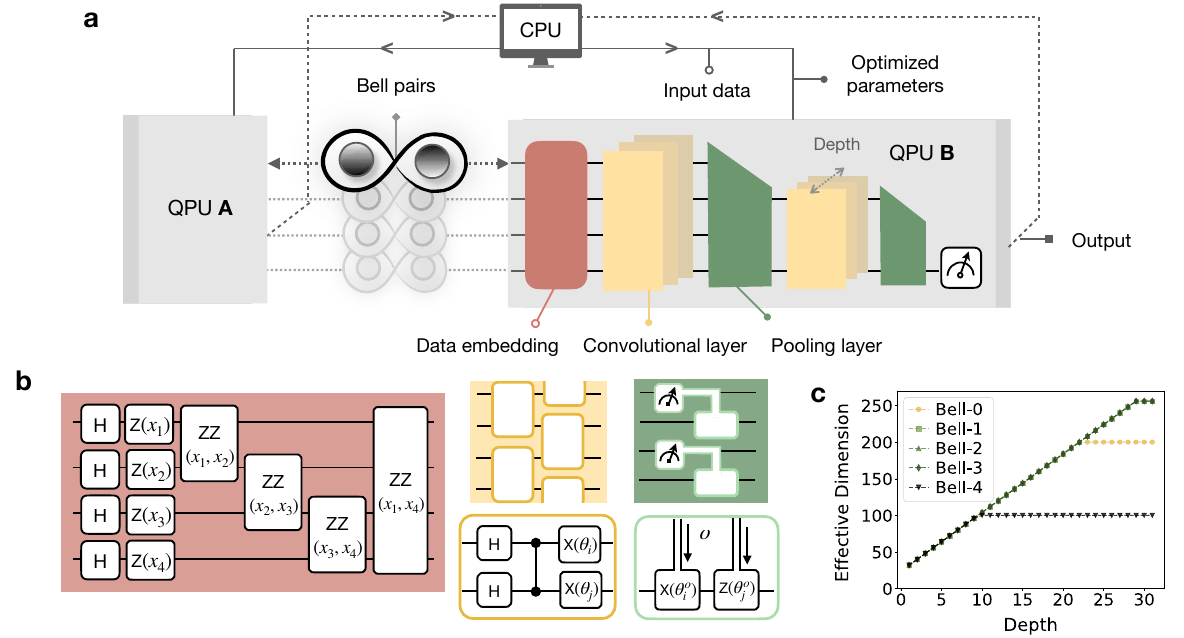}

\caption{\textbf{Distributed quantum machine learning with pre-shared entanglement.} 
\textbf{a}, Pre-shared Bell pairs connect the two quantum processors. Partitioned input data are embedded locally and processed through convolutional and pooling layers. The outputs are then used to update the training parameters or to perform classification.
\textbf{b}, (red) Input data $\mathbf{s}$ and $\mathbf{t}$ are mapped to gate parameters $\mathbf{x}$ and embedded using $Z$ rotations with angles $x_i$ and $ZZ$ interactions with angles $\frac{1}{2}(\pi - x_i)(\pi - x_j)$. 
(yellow) Convolutional layers consist of building blocks arranged in a brick-wall pattern, with variational parameters assigned by qubit and depth. 
(green) Pooling layers apply measurement-conditioned operations to adjacent qubits.
\textbf{c}, The effective dimension as a function of convolutional-layer depth. The data-embedding layer is replaced with Haar-random local unitaries. The maximum effective dimension as a function of the number of Bell pairs is $200$, $256$, $256$, $256$, and $100$, respectively.
}
\label{fig2}
\end{figure*}

Unlike direct communication, entanglement can be established offline~\cite{azuma2023quantum}, enabling purification to high fidelity~\cite{bennett1996purification} and long-term storage~\cite{lei2023quantum,wang2021single,wang2025nuclear}. Entanglement has been distributed over hundreds of kilometers via satellite links and optical fiber networks, achieving fidelities on the order of 85\%~\cite{yin2017satellite,wengerowsky2020passively,neumann2022continuous}, while laboratory experiments have reached fidelities of up to 97\%~\cite{saha2025highfidelity}. Although entanglement alone cannot be used for signaling, it can reduce communication costs or enhance success probabilities in distributed tasks~\cite{brassard2003quantum,brukner2004bells,buhrman2010nonlocality}. This advantage is exemplified by the CHSH game, where two spatially separated parties must produce correlated outputs without communication. While classical strategies are limited to a maximum success probability of 75\%, pre-shared entanglement increases this to approximately 85\%~\cite{CHSH1969}. Such advantages in communication complexity problems (CCPs) can generally be achieved using nonlocal correlations that violate Bell-like inequalities~\cite{brunner2014bell,buhrman2016quantum,ho2022entanglement}.

Distributed quantum machine learning (DQML) exhibits structural similarities with CCPs, where spatially separated processors learn from local inputs under restricted communication (Fig.~\ref{fig1}\textbf{b})~\cite{hwang2025DQML}. This analogy suggests that shared entanglement can improve machine-learning performance. While entanglement has been identified as a key resource for quantum advantage in machine learning, enabling enhanced expressivity and inductive biases beyond classical models~\cite{gao2022enhancing,anschuetz2023interpretable,bowles2023contextuality}, its role in distributed settings from a communication complexity perspective remains largely unexplored.

Here, we numerically investigate the utility of pre-shared entanglement in DQML using PennyLane~\cite{bergholm2018pennylane}, considering binary classification tasks with two processors and varying numbers of shared Bell pairs. We begin with a dataset inspired by the extended CHSH game~\cite{brukner2004bells}, and show that a loss function based on the CHSH correlator is highly sensitive to data embedding, whereas the commonly used mean-squared-error loss exhibits greater robustness. We then evaluate performance on synthetic datasets and find that even a single Bell pair consistently improves classification accuracy. Remarkably, performance deteriorates under maximal entanglement, although this degradation can be mitigated by optimizing the entanglement structure. Our results identify entanglement as a key communication resource and provide a practical route to DQML beyond coherence-time limitations.\\

\begin{figure*}
\centering
\includegraphics[width=1\linewidth]{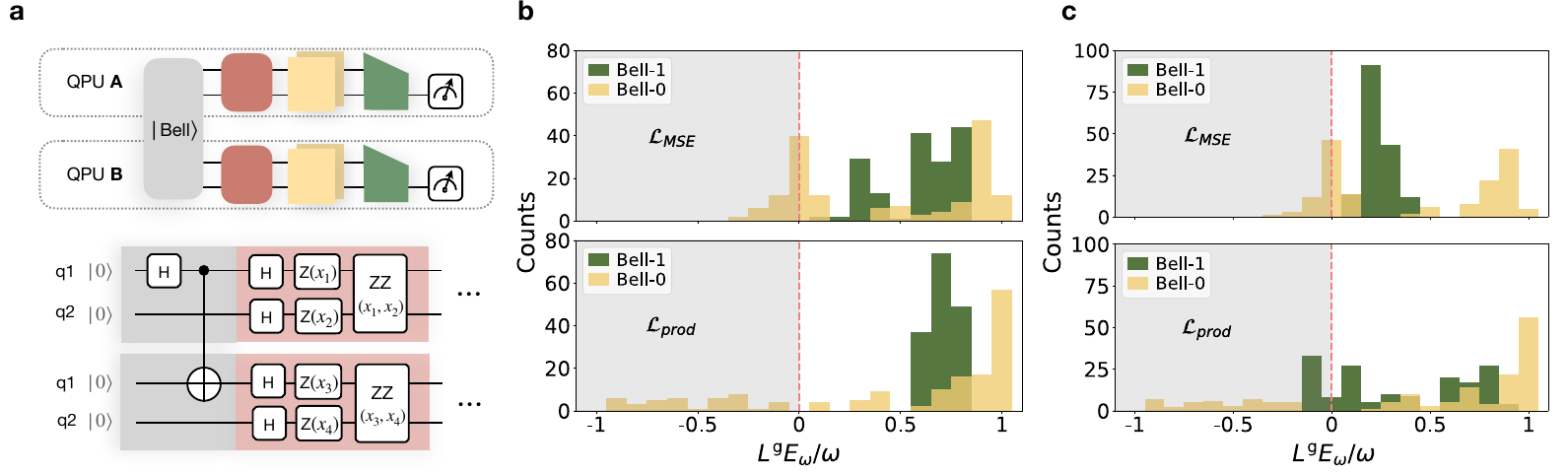}
\caption{\textbf{DQML for extended CHSH dataset.} \textbf{a}, The CHSH-inspired length-4 dataset is partitioned between two processors, each consisting of two qubits, with and without a single Bell pair (Bell-1 and Bell-0). The convolutional and pooling layers use the same circuit blocks as in Fig.~\ref{fig2}\textbf{b}. The convolutional-layer depth is fixed at 20, and the weight vector $\boldsymbol{\omega}$ is constrained to the parity form.
\textbf{b}, Histograms of $L^\mathrm{g} E_{\boldsymbol{\omega}}/\omega$ over all 16 inputs, aggregated over ten independent training runs using the optimal data embedding $\mathbf{x}^o$. Both the MSE and product loss functions achieve 100\% classification accuracy for Bell-1 and 75\% for Bell-0.
\textbf{c}, Histograms obtained using the alternative data embedding $\mathbf{x}^a$. The MSE loss preserves the accuracy obtained with the optimal embedding, whereas the product loss achieves 75\% classification accuracy for both Bell-1 and Bell-0. }
\label{fig3}
\end{figure*}

\noindent{\fontsize{11.2}{13}\selectfont\textbf{Results}}\vspace{0.3em}\\ 
\noindent\textbf{Long-distance DQML schematic.} As illustrated in Fig.~\ref{fig2}\textbf{a}, two four-qubit processors are linked by pre-shared Bell pairs. Input data is partitioned into $\mathbf{s}$ and $\mathbf{t}$, 
each processed locally to produce one-bit outcomes $a$ and $b$. The dataset provides the ground-truth label $L^{\mathrm{g}}(\mathbf{s},\mathbf{t})$, and a trained classifier infers the predicted label $L^{\mathrm{p}}(\mathbf{s},\mathbf{t})$ from these outcomes. Rather than using a deterministic strategy, we employ a probabilistic strategy that leverages the full output distribution to enhance prediction performance. Accordingly, the classifier is constructed from a weighted sum of the joint outcome probabilities,
\begin{equation}
    E_{\boldsymbol{\omega}}(\mathbf{s},\mathbf{t})
    = \sum_{a,b\in\{0,1\}} \omega_{ab}\, P(a,b|\mathbf{s},\mathbf{t}),
    \label{eq:expectation}
\end{equation}
where $P(a,b|\mathbf{s},\mathbf{t})$ is the joint probability and $\omega_{ab}$ denotes the corresponding weight. In this work, the predicted label is determined as the sign of the value,
\begin{equation}
L^{\mathrm{p}}(\mathbf{s},\mathbf{t}) 
= \mathrm{sgn}\!\left[E_{\boldsymbol{\omega}}(\mathbf{s},\mathbf{t})\right] \in \{-1,+1\}.
\label{eq:L_pred}
\end{equation}

We now describe the data-embedding and training circuits used in our DQML framework (Figs.~\ref{fig2}\textbf{a},\textbf{b}). The two processors, with $n$ pre-shared Bell pairs, are initialized in
\begin{equation}
|\mathrm{Bell}\text{-}n\rangle = \bigotimes_{j=1}^n |\Phi^+\rangle_{A_jB_j}
\bigotimes_{k=n+1}^{4} |00\rangle_{A_kB_k},
\label{init_state}
\end{equation}
where $|\Phi^+\rangle = (|00\rangle + |11\rangle)/\sqrt{2}$, and the subscript $j$ denotes the $j$-th qubit pair across processors $A$ and $B$. The input data $\mathbf{s}$ and $\mathbf{t}$ are then encoded into the gate parameters $\mathbf{x}$ of the embedding layer (red)~\cite{havlivcek2019supervised}, such that the pre-shared entanglement effectively forms part of the data embedding. The state is subsequently processed by a quantum convolutional neural network (QCNN)~\cite{cong2019qcnn}, which consists of alternating convolutional and pooling layers. The convolutional layers are arranged in a brick-wall pattern using the yellow building block repeated over multiple depths, while the pooling layers employ the green building block, which applies measurement-conditioned operations on adjacent qubits to produce a single binary output per processor. Further details of the training circuit are provided in Supplementary Note~1.

To construct an optimal classifier, we jointly optimize the circuit parameters and the weight vector $\boldsymbol{\omega}$ in Eq.~\eqref{eq:expectation} by minimizing a loss function over a training dataset. Within the CCP framework, this corresponds to optimizing the observable $\hat{\mathcal{O}}_{\boldsymbol{\omega}} = \sum_{a,b} \omega_{ab} \, |ab\rangle \langle ab|,$ which is effectively transformed by the data-embedding, convolutional, and pooling layers. In this picture, the input data specify the observable, while the trainable circuit parameters rotate the observable to maximize correlations with the ground-truth label. Just as Bell inequality violations depend on the choice of measurement observables, so does the benefit of entanglement in communication complexity problems. Further details on the CCP analogy are provided in Supplementary Note~2.

To probe how circuit expressivity depends on the number of pre-shared Bell pairs (Eq.~\eqref{init_state}), we estimate the effective dimension as a function of convolutional-layer depth, using input states generated by replacing the data-embedding layer with Haar-random local unitaries (see Methods for details). The effective dimension quantifies the number of trainable parameters that meaningfully contribute to learning~\cite{abbas2021qnn,haug2021capacity}. Figure~\ref{fig2}\textbf{c} shows that all cases exhibit similar growth at shallow depths. However, as the depth increases, configurations with intermediate entanglement achieve the largest effective dimension, whereas the maximally entangled case (Bell-4) saturates early and yields the smallest value, even below that of the separable case (Bell-0). This implies that excessive entanglement may degrade DQML performance, 
consistent with prior studies on the computational power of highly entangled states~\cite{gross2009tooentangled,hayden2006generic,bremner2009randomstates}.\\

\begin{figure*}
\centering
\includegraphics[width=0.78\linewidth]{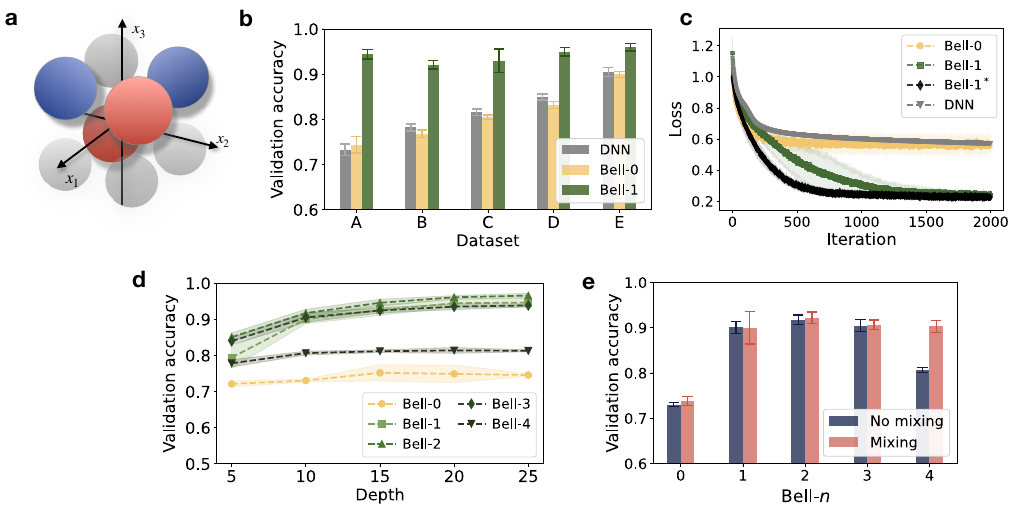}
\caption{\textbf{DQML for synthetic datasets.}
\textbf{a}, Visualization of a simplified three-dimensional synthetic dataset, where data points $\mathbf{x}$ form two distinct clusters, with red and blue balls denoting ground-truth labels $+1$ and $-1$, respectively. For the DQML tasks, datasets are constructed with the same method in eight-dimensional space. 
\textbf{b}, Validation accuracy across five eight-dimensional datasets at a convolutional-layer depth of 20 after 2000 training iterations. DNN denotes the result obtained with a distributed classical neural network 
using a comparable number of trainable parameters to the Bell-$n$ configurations. The following panels show results obtained on dataset A with the same training iterations. 
\textbf{c}, Training loss as a function of iteration, where Bell-1$^*$ denotes the configuration in which the weight vector $\boldsymbol{\omega}$ is fixed in parity form. 
\textbf{d}, Validation accuracy as a function of convolutional-layer depth. 
\textbf{e}, Validation accuracy with and without entanglement mixing layers, where the mixing configuration uses 3 trainable layers prior to data embedding and 7 convolutional layers, and the no-mixing configuration uses 10 convolutional layers. Error bars and shaded regions indicate one standard deviation over ten trials for \textbf{b,c,d} and five for \textbf{e}.}
\label{fig4}
\end{figure*}

\noindent\textbf{Extended CHSH dataset.} We first benchmark DQML on a length-4 dataset, $(s_1,s_2,t_1,t_2)$, inspired by the extended CHSH game, where entanglement advantage in communication complexity is well established~\cite{brukner2004bells}. In the task, the ground-truth label is defined as
\begin{equation}
L^\mathrm{g}(\mathbf{s}, \mathbf{t}) = s_2 t_2 (-1)^{s_1 t_1},
\label{eq:CHSH}
\end{equation}
for $s_1,t_1 \in \{0,1\}$ and $s_2,t_2 \in \{-1,1\}$. The partitioned inputs $\mathbf{s}=(s_1,s_2)$ and $\mathbf{t}=(t_1,t_2)$ are embedded into each two-qubit processor as shown in Fig.~\ref{fig3}\textbf{a}. Under the optimal embedding $\mathbf{x}^{o} = \frac{\pi}{2}(s_1,s_2,t_1,t_2)$, the bits $s_1$ and $t_1$ directly transform the Bell state, effectively selecting local measurement bases with an angular separation of $\pi/2$ on the Bloch sphere. The alternative embedding $\mathbf{x}^{a} = \frac{\pi}{2}(s_2,s_1,t_2,t_1)$, by contrast, induces a separation of $\pi$. Since saturating Tsirelson's bound requires the local observables to differ by $\pi/2$~\cite{cirel1980quantum}, $\mathbf{x}^{o}$ is expected to yield greater advantage in our distributed task. See Supplementary Note~3 for details.

For training, we consider two loss functions: the product loss and the mean-squared-error (MSE) loss. The product loss is motivated by the CHSH correlator $S = \langle A_0B_0\rangle + \langle A_0B_1\rangle + \langle A_1B_0\rangle - \langle A_1B_1\rangle$, where $\langle A_jB_k\rangle$ denotes the expectation value of the joint 
observable determined by inputs $j$ and $k$. The loss function is 
\begin{equation}
\mathcal{L}_{\text{prod}} = -\frac{1}{N} \sum_{\mathbf{s},\mathbf{t}} L^{\mathrm{g}}(\mathbf{s},\mathbf{t}) \, E_{\boldsymbol{\omega}}(\mathbf{s},\mathbf{t}),
\label{eq:Loss_prod}
\end{equation}
where $N$ is the number of training samples. With the label in Eq.~\eqref{eq:CHSH}, this loss is analogous to the CHSH correlator for  inputs $s_1$ and $t_1$. The MSE loss is defined as
\begin{equation}
\mathcal{L}_{\text{MSE}} = \frac{1}{N} \sum_{\mathbf{s},\mathbf{t}} \left| L^{\mathrm{g}}(\mathbf{s},\mathbf{t}) - E_{\boldsymbol{\omega}}(\mathbf{s},\mathbf{t}) \right|^2.
\label{eq:Loss_MSE}
\end{equation}

We evaluate classification performance using histograms of $L^{\mathrm{g}} E_{\boldsymbol{\omega}}/\omega$ over all 16 inputs across ten trials. Since the predicted label is $L^\mathrm{p}=\mathrm{sgn}[E_{\boldsymbol{\omega}}]$ (Eq.~\eqref{eq:L_pred}), positive values indicate correct classification. We set the weight vector to the parity form $\boldsymbol{\omega} = (\omega, -\omega, -\omega, \omega)$ for both loss functions, as the parity measurement is known to be optimal for the CHSH game. Here $\omega > 0$ is treated as a trainable parameter for the MSE loss, but fixed to unity for the product loss to avoid trivial rescaling.

Under the optimal embedding $\mathbf{x}^o$, both loss functions achieve 100\% classification accuracy for Bell-1 and 75\% for Bell-0 with 20 convolutional layers (Fig.~\ref{fig3}\textbf{b}). The normalized quantity $L^{\mathrm{g}} E_{\boldsymbol{\omega}}/\omega\in[-1,1]$ can be converted to the CHSH success probability for each input through $ p_{\text{succ}}=\frac{1}{2}(L^{\mathrm{g}} E_{\boldsymbol{\omega}}/\omega +1)$. For the product loss, corresponding to CHSH correlator optimization, the average success probabilities are  found to be 85.2\% for Bell-1 and 74.4\% for Bell-0, in close agreement with the quantum (85.3\%) and classical (75\%) bounds~\cite{buhrman2010nonlocality}. This confirms that training reaches the fundamental bounds, and that the entanglement advantage in DQML shares the same origin as in the CHSH game.

Under the alternative embedding $\mathbf{x}^a$, the product loss achieves 75\% classification accuracy for both Bell-1 and Bell-0 (Fig.~\ref{fig3}\textbf{c}). This reflects the fact that Tsirelson’s bound for the CHSH correlator $S$ is attained only under optimal measurement settings. By contrast, the MSE loss preserves the accuracy obtained with the optimal embedding, highlighting its robustness to suboptimal embeddings. The degradation of the product loss can be attributed to its  structure in Eq.~\eqref{eq:Loss_prod}: large values of $L^\mathrm{g}E_{\boldsymbol{\omega}}$ on a subset of inputs can dominate the objective even at the cost of misclassifying others, whereas the MSE loss penalizes each input independently and enforces more balanced optimization.\\

\noindent\textbf{Synthetic dataset.} We next consider a binary classification task on synthetic datasets using two four-qubit processors, employing the embedding and training circuit shown in Fig.~\ref{fig2}. Because commonly used datasets such as MNIST and Iris are readily trainable with a single four-qubit processor~\cite{abbas2021qnn,hur2022quantum}, we instead generate clustered eight-dimensional datasets. The ground-truth binary labels are randomly assigned to 64 clusters, and each length-8 input vector is partitioned into two length-4 subsets for separate embedding, $\mathbf{x}=(\mathbf{s},\mathbf{t})$ (see Fig.~\ref{fig4}\textbf{a} for a simplified example). We use three quarters of the data for training with the MSE loss and the remaining quarter for validation. The weight vector $\boldsymbol{\omega}$ is optimized without any constraints. Further details are provided in Methods. As a baseline, we introduce a distributed classical neural network (DNN)~\cite{sheela2013review} that processes the partitioned data locally and determines the label from one-bit outputs at each node (see Supplementary Note~4 for details). 

Figure~\ref{fig4}\textbf{b} showcases the classification accuracy across five datasets with 20 convolutional layers after 2000 training iterations, showing that even a single Bell pair yields a substantial improvement over the no-communication baselines, particularly for the more challenging datasets. The training loss on dataset A in Fig.~\ref{fig4}\textbf{c} highlights this advantage from a complementary perspective. Bell-1 and Bell-1$^*$ both converge to significantly lower loss values than Bell-0 and DNN. For Bell-1$^*$, the weight vector is fixed in the parity form during training. Although unconstrained optimization of Bell-1 introduces trial-to-trial variation, Bell-1$^*$ consistently achieves fast convergence, indicating that the parity measurement is near-optimal for this classification task, in line with a recent result showing that product measurements can achieve optimal entanglement-assisted performance~\cite{piveteau2022entanglementsimple}. We also observe that, despite having a comparable number of trainable parameters, Bell-0 converges faster than DNN, demonstrating that quantum machine learning provides better training efficiency than its classical counterpart~\cite{hur2022quantum}.

Figure~\ref{fig4}\textbf{d} presents the classification accuracy on dataset A as a function of convolutional-layer depth for different numbers of Bell pairs. For Bell-1,2,3, performance is similar across configurations, consistently exceeding 90\% accuracy from a depth of 15 onward. The maximally entangled Bell-4, by contrast, saturates near 80\% even at large circuit depths, consistent with its reduced maximum effective dimension (Fig.~\ref{fig2}\textbf{c}). Notably, despite its smaller effective dimension relative to Bell-0, Bell-4 still outperforms the separable case, indicating that nonlocal correlations confer an advantage beyond what is captured by circuit expressivity alone.

To further probe the role of entanglement structure, we introduce trainable entanglement mixing layers prior to data embedding (see Supplementary Note~5 for details). The mixing layers allow diverse entanglement structures to be explored within a fixed entanglement budget, such as an 8-qubit GHZ state generated from a single Bell pair.  As shown in Fig.~\ref{fig4}\textbf{e}, this restructuring enables Bell-4 to recover classification accuracy comparable to intermediate entanglement levels. Since the effective dimension of Bell-4 does not increase upon restructuring (Supplementary Note~5), the accuracy improvement cannot be attributed to enhanced circuit expressivity. These results highlight that entanglement structure, rather than entanglement quantity or effective dimension themselves, is the key factor governing learning performance in DQML.\\

\noindent{\fontsize{11.2}{13}\selectfont\textbf{Discussion}}\vspace{0.3em}\\
Our results have established pre-shared entanglement as a genuine communication resource in DQML. The agreement between trained classifiers and the theoretical bounds of the CHSH game has confirmed that gradient-based optimization reliably discovers entanglement-assisted communication advantages. We have also found that the standard MSE loss is robust to suboptimal data embeddings, in contrast to the product loss inspired by the CHSH correlator. Extending this to eight-dimensional datasets, we have observed that even a single Bell pair yields accuracy improvements of up to 20\% over the no-communication baselines. 

An intriguing finding is the non-monotonic dependence of classification performance on entanglement quantity. While Bell-1,2,3 consistently achieve accuracies exceeding 90\%, the maximally entangled Bell-4 configuration saturates near 80\%, consistent with its reduced effective dimension. Crucially, introducing trainable layers that restructure the entanglement prior to data embedding has recovered this performance gap, even without increasing the effective dimension. This indicates that the learning advantage is rooted in task-relevant nonlocal correlations rather than parametric circuit expressivity. 

These findings carry direct implications for future quantum networks. Because our protocols rely on pre-shared entanglement rather than real-time communication, they are inherently compatible with global-scale networks where propagation delays exceed qubit coherence times. Taken together, our results suggest that the key design principle for DQML is not the maximization of entanglement, but its careful structuring to match the geometry of the learning task. Extending this framework to many-node networks and realistic noise models are natural next steps toward practical quantum internet-enabled machine learning.\\

\noindent{\fontsize{11.2}{13}\selectfont\textbf{Methods}}\vspace{0.3em}\\
\noindent\textbf{Effective dimension.}
\noindent We quantify circuit capacity using a rank-based effective dimension, defined as the maximal rank of the Fisher information matrix over the parameter space $\boldsymbol{\theta} \in \boldsymbol{\Theta}$~\cite{abbas2021qnn,haug2021capacity}:
\begin{equation}
\mathrm{ED} = \max_{\boldsymbol{\theta} \in \boldsymbol{\Theta}} \mathrm{rank}\, F(\boldsymbol{\theta}).
\end{equation}
For a given parameter setting $\boldsymbol{\theta}$, the Fisher information matrix averaged over inputs is defined as
\begin{equation}
F_{ij}(\boldsymbol{\theta})
= \frac{1}{N}\sum_{n=1}^{N}\sum_{\boldsymbol{y}}
P(\boldsymbol{y}|\boldsymbol{x}_n;\boldsymbol{\theta})\,
\frac{\partial \log P}{\partial \theta_i}
\frac{\partial \log P}{\partial \theta_j},
\end{equation}
where $P(\boldsymbol{y}|\boldsymbol{x}_n;\boldsymbol{\theta})$ denotes the conditional probability distribution over computational basis outputs $\boldsymbol{y}$ for the $n$-th input $\boldsymbol{x}_n$, and $N$ is the total number of data points. The rank of $F(\boldsymbol{\theta})$ characterizes the number of locally independent parameter directions affecting the model likelihood. 

To characterize the general expressivity of the training circuit, we numerically estimate the effective dimension over 100 Haar-random datasets, each constructed by embedding the input data using a unitary drawn from the Haar measure. For each dataset, the Fisher information matrix is evaluated at 20 parameter points sampled from a uniform distribution, and the maximum observed rank is reported in Fig.~\ref{fig2}\textbf{d}.\\

\noindent\textbf{Synthetic data classification.}
\noindent We generate synthetic eight-dimensional datasets following the procedure described in \cite{hwang2025DQML}. We first sample 4096 points $\mathbf{x} = (x_1,\ldots,x_8)$ uniformly from a ball of radius $\pi/4$, defined by $\sum_{i=1}^{8} x_i^2 \le (\pi/4)^2$. The samples are randomly grouped into 64 clusters, each assigned a binary label $L \in \{-1, +1\}$, with the number of samples in each class kept balanced. To generate well-separated clusters in the eight-dimensional space, each sample is shifted according to $\mathbf{x} \rightarrow \mathbf{x} + \mathbf{v}_L$, where the label-dependent shift vector $\mathbf{v}_L = (v_1,\ldots,v_8)$ has components $v_i \in \{-\pi/4,\pi/4\}$. The resulting dataset is split into training and validation sets in a 3:1 ratio. For all results in Fig.~\ref{fig4}, training is performed for 2000 iterations with a batch size of one quarter of the training set.\\

\noindent{\fontsize{11.2}{13}\selectfont\textbf{Data availability}}

\noindent The data that support the findings of this study are available from the corresponding author upon request.\\

\noindent{\fontsize{11.2}{13}\selectfont\textbf{Code availability}}

\noindent The code used to generate the figures within this paper and other findings of this study are available from the corresponding author upon request.

\bibliographystyle{naturemag}
\bibliography{Bib}
\vspace*{0.5em}

\newpage

\noindent{\fontsize{11.2}{13}\selectfont\textbf{Acknowledgments}} 

\noindent The authors thank Jiwon Yune and Eunsung Kim for their thoughtful discussions. This work was partly supported by National Research Foundation of Korea (NRF) grant funded by the Korea government (MSIT) (RS-2024-00353348, RS-2024-00432563, RS-2024-00404854, RS-2024-00438415, RS-2025-25464760) and a Korea University Grant. H.~K. is supported by the KIAS Individual Grant No. CG085302 at Korea Institute for Advanced Study.\\

\noindent{\fontsize{11.2}{13}\selectfont\textbf{Author contributions}}

\noindent Yosep K. initiated the project. Yosep K. and H. K. supervised the research. Yerim K. performed the numerical experiments and carried out the theoretical analysis with input from K. H. Yerim K. and Yosep K. wrote the manuscript with input from all authors.\\

\noindent{\fontsize{11.2}{13}\selectfont\textbf{Competing interests}}

\noindent The authors declare no competing interests.\\

\noindent{\fontsize{11.2}{13}\selectfont\textbf{Additional information}}

\noindent Correspondence and requests for materials should be addressed to Yosep Kim.

\newpage
\clearpage

\renewcommand{\theequation}{S\arabic{equation}}
\renewcommand{\thefigure}{S\arabic{figure}}
\renewcommand{\thetable}{S\arabic{table}}

\setcounter{page}{1}
\setcounter{figure}{0}
\setcounter{table}{0}
\setcounter{equation}{0}

\onecolumngrid
\vspace{\columnsep}
\begin{center}
\large \textbf{Supplementary Materials for \\ The power of entanglement in distributed quantum machine learning}
\end{center}

\input{supplementary}

\end{document}

%% file: Supplementary.tex
\section{Supplementary Note 1 --- Details on the Training Circuit}

\begin{figure}[b]
\centering
\includegraphics[width=0.58\linewidth]{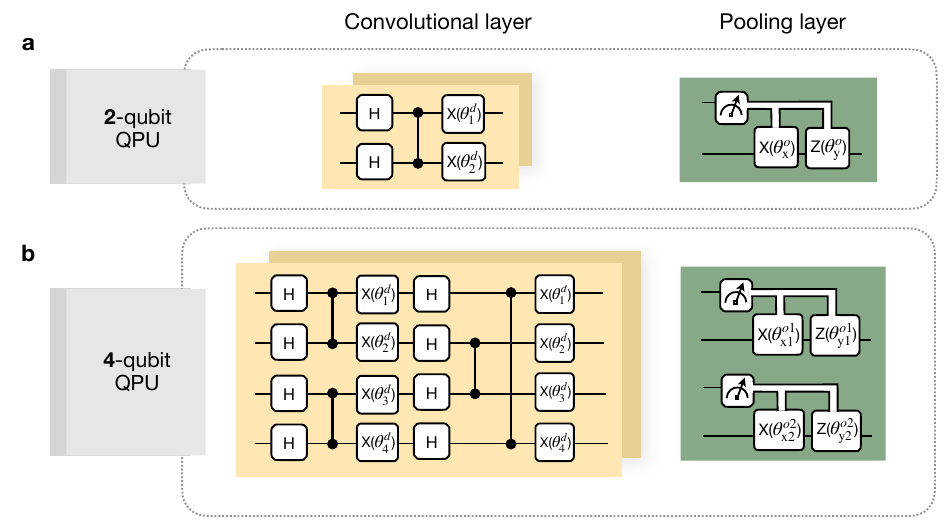}
\caption{\textbf{Details of the convolutional and pooling layers.}
The convolutional layer (yellow) contains two trainable parameters per depth in the 2-qubit circuit and four in the 4-qubit circuit. The pooling layer (green) applies measurement-conditioned X and Z rotations on the remaining qubits, controlled by the measurement outcomes of the measured qubits, introducing four trainable parameters in the 2-qubit circuit and eight in the 4-qubit circuit.}
\label{details_circuit}
\end{figure}

\begin{figure}[hb!]
\centering
\includegraphics[width=0.5\linewidth]{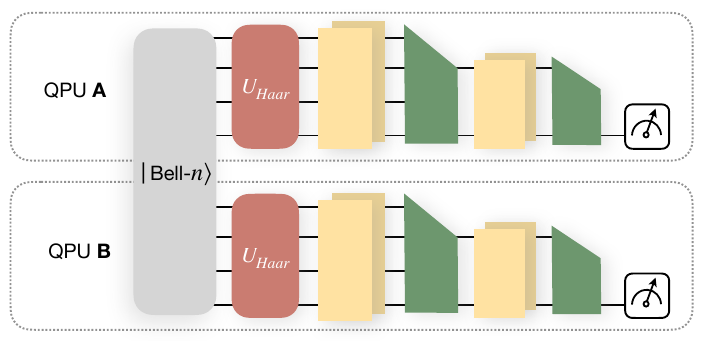}
\caption{\textbf{Circuit used for effective dimension estimation.}
Haar-random unitaries $U_\mathrm{Haar}$ replace the data-embedding layers 
of each QPU, with the remaining circuit structure unchanged.}
\label{effectdim_circ}
\end{figure}

The quantum circuits used for distributed quantum machine learning (DQML) are detailed in Fig.~\ref{details_circuit}, which shows the convolutional and pooling blocks for a single circuit depth in the 2-qubit and 4-qubit settings. The convolutional layers are arranged in a brick-wall pattern, where each two-qubit block consists of Hadamard gates followed by a CZ gate and single-qubit X rotations. In the convolutional block, the trainable parameters are denoted by $\theta_i^d$, where $d$ labels the circuit depth and $i$ indexes the qubit on which the rotation is applied; each depth contains two trainable parameters in the 2-qubit circuit and four in the 4-qubit circuit. 

The pooling block uses measurement-conditioned rotations to reduce the number of active qubits. In the 2-qubit circuit, the measurement outcome $o \in \{0,1\}$ of the first qubit classically controls the $X(\theta_\mathrm{x}^o)$ and $Z(\theta_\mathrm{y}^o)$ gates applied to the remaining qubit, introducing four trainable parameters in total. In the 4-qubit circuit, the first and third qubits are measured, and their outcomes $o_1, o_2 \in \{0,1\}$ control the X and Z rotations on the second and fourth qubits, respectively, introducing eight trainable parameters in total. After one pooling block, the 4-qubit circuit is reduced to a 2-qubit architecture, after which the subsequent layers follow the same structure as in the 2-qubit case.

To estimate the effective dimension shown in Fig.~2\textbf{c} of the main text, we use the circuit illustrated in Fig.~\ref{effectdim_circ}, in which Haar-random unitaries replace the data-embedding layers of each processor. The convolutional-layer depth is increased incrementally, and the effective dimension is evaluated until no further increase is observed. For each parameter set, the Fisher information matrix is averaged over 100 Haar-random unitaries, and this procedure is repeated for 20 random parameter sets. The effective dimension is then taken as the maximum observed rank.

\begin{figure}[b]
\centering
\includegraphics[width=0.54\linewidth]{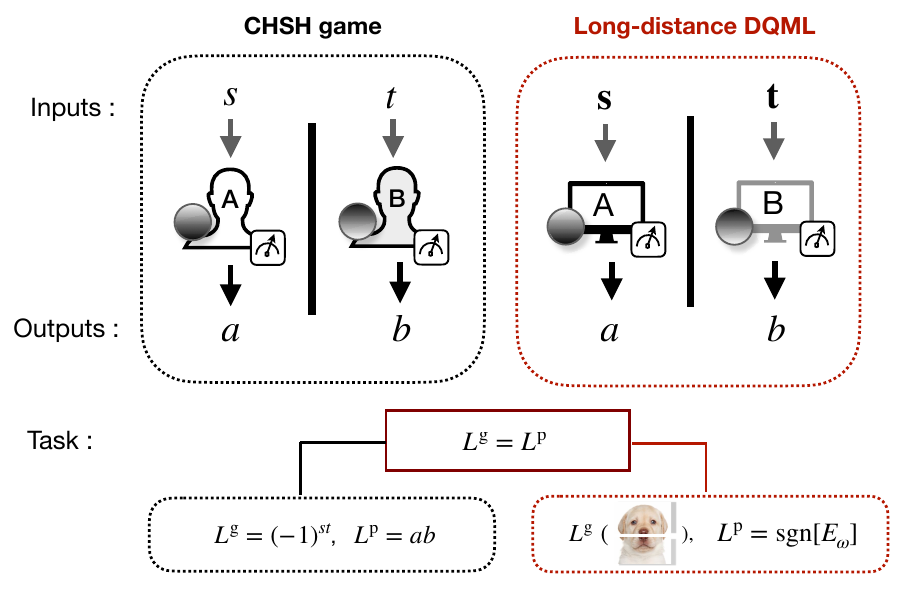}
\caption{\textbf{Analogy between the CHSH game and long-distance DQML.} 
Both tasks require two spatially separated parties to evaluate a global binary-valued function $L^{\mathrm{g}}$ from local inputs without communication. In the CHSH game, $L^{\mathrm{p}} = ab$ is assigned from a single-shot outcome and performance is characterized by success probability. In DQML, $L^{\mathrm{p}} = \mathrm{sgn}[E_{\boldsymbol{\omega}}]$ is inferred from repeated measurements and performance is characterized by classification accuracy.}
\label{ccp}
\end{figure}

\section{Supplementary Note 2 --- Analogy with Communication Complexity Problems}

Communication complexity problems (CCPs) require spatially separated parties to evaluate a global function from local inputs $s$ and $t$ without communication. The task succeeds if Alice and Bob reproduce the target value through locally generated outputs $a,b\in\{-1,1\}$. In the absence of communication, nonlocal correlations provided by pre-shared entanglement can improve the success probability~\cite{buhrman2010nonlocality}. When Alice and Bob are replaced by remote quantum processors, the CCP setting becomes structurally equivalent to DQML, where the global function is identified with the ground-truth label $L^{\mathrm{g}}$ and the local outputs define the predicted label $L^{\mathrm{p}}$ (Fig.~\ref{ccp}). For binary-valued targets, the problem can thus be interpreted as a binary classification task.

In the CHSH game, the predicted label is determined from a single-shot measurement outcome as the parity $L^{\mathrm{p}} = ab$~\cite{CHSH1969}. In our DQML protocol, by contrast, $L^{\mathrm{p}}$ is inferred from the expectation value of the joint output distribution over repeated measurements,
\begin{equation}
E_{\boldsymbol{\omega}}(\mathbf{s},\mathbf{t})
= \sum_{a,b\in\{-1,1\}} \omega_{ab}\, P(a,b|\mathbf{s},\mathbf{t}),
\end{equation}
and defined as its sign, $L^{\mathrm{p}} = \mathrm{sgn}[E_{\boldsymbol{\omega}}]$ where $\boldsymbol{\omega}=(1,-1,-1,1)$.

This distinction leads to different notions of success in the two settings. In DQML, the classification accuracy is defined as the fraction of instances for which the predicted label matches the ground-truth label,
\begin{equation}
\mathcal{A}_{\mathrm{class}} = \frac{1}{N}\sum_{i=1}^{N} 
\mathbf{1}\left[L_i^{\mathrm{p}} = L_i^{\mathrm{g}}\right],
\label{supp:DQMLacc}
\end{equation}
where $\mathbf{1}[\cdot]$ denotes the indicator function, which equals 1 if the condition in the bracket is satisfied and 0 otherwise, and $N$ is the total number of samples. In the CHSH game, the success probability for a given input is
\begin{equation}
\mathcal{P}_{\mathrm{succ}} = \frac{1 + L^{\mathrm{g}} E_{\boldsymbol{\omega}}}{2},
\label{CHSHsuccessprob}
\end{equation}
which is sensitive to the magnitude of $E_{\boldsymbol{\omega}}$, not only its sign. Accordingly, the DQML classification accuracy is governed by the fraction of samples with the correct sign of $E_{\boldsymbol{\omega}}$, whereas the CHSH success probability also depends on how far the distribution is shifted from zero. In Fig.~3 of the main text, the classification accuracy is determined by the fraction of positive samples, whereas the CHSH success probability additionally reflects the magnitude of the shift.
\\

\begin{table}[t]
\centering
\begin{tabular}{c|ccc}
\hline
 & \textbf{CHSH} & \textbf{Extended CHSH} & \textbf{Domain} \\
\hline
\textbf{Input} 
& $\{s_1,\, t_1\}$ 
& $\{s_1,\, s_2,\, t_1,\, t_2\}$ 
& $s_1,t_1, \in \{0,1\},\quad s_2,t_2 \in\{-1,1\}  $\\[3pt]

\textbf{Output} 
& $\{a,\, b\}$ 
& $\{a,\, b\}$ 
& $a,b\in\{-1,1\}$ \\[3pt]

$\mathbf{L}$ 
& $(-1)^{s_1  t_1}$ 
& $ s_2t_2(-1)^{s_1t_1}$ 
& $L\in\{-1,1\}$ \\[3pt]

$\mathbf{f}$ 
& $ab$ 
& $ab$ 
& $f\in\{-1,1\}$ \\
\hline
\end{tabular}
\caption{\textbf{Comparison of the standard and extended CHSH games.}
The table summarizes the inputs, outputs, binary-valued target functions, and prediction functions used in the two settings.}
\label{tab:chsh_extended}
\end{table}

\begin{table}[t]
\centering
\begin{tabular}{c c c | c c c}
\hline
Run & Bell-1 & Bell-0 & Run & Bell-1 & Bell-0 \\
\hline
1  & 0.8525 & 0.7407 & 6  & 0.8511 & 0.7408 \\
2  & 0.8519 & 0.7412 & 7  & 0.8518 & 0.7364 \\
3  & 0.8514 & 0.7437 & 8  & 0.8520 & 0.7412 \\
4  & 0.8524 & 0.7406 & 9  & 0.8525 & 0.7409 \\
5  & 0.8512 & 0.7438 & 10 & 0.8517 & 0.7410 \\
\hline
\end{tabular}
\caption{\textbf{CHSH success probabilities.}
CHSH success probabilities for the Bell-1 and Bell-0 scenarios over 10 independent trials, obtained using the product loss and the optimal embedding $\mathbf{x}^o$. The product loss is constructed from the CHSH correlator $S = \langle A_0B_0\rangle + \langle A_0B_1\rangle + \langle A_1B_0\rangle - \langle A_1B_1\rangle$. Bell-1 approaches the quantum bound while Bell-0 approaches the classical bound, consistent with the optimal CHSH strategy.}
\label{tab:extended_tsirelson}
\end{table}

\section{Supplementary Note 3 - Extended CHSH game}

In the extended CHSH game, the global function, or equivalently the ground-truth label, is defined as $L^{\mathrm{g}}(\mathbf{s},\mathbf{t}) = s_2 t_2 (-1)^{s_1 t_1}$~\cite{brukner2004bells}. The task is to make the predicted label $L^{\mathrm{p}} = ab$ coincide with the ground-truth label, i.e., $L^{\mathrm{p}} = L^{\mathrm{g}}$.
Writing the two parties' outputs as $a = s_2\alpha$ and $b = t_2\beta$, this condition reduces to $\alpha \beta = (-1)^{s_1 t_1}$, which is precisely the winning condition of the standard CHSH game. Thus, the optimal strategy is inherited directly from the CHSH protocol, with $\alpha$ and $\beta$ chosen to saturate Tsirelson's bound~\cite{cirel1980quantum,buhrman2010nonlocality}.

Equivalently, the task can be expressed in terms of the CHSH correlator
\begin{equation}
    S =
    \langle A_0B_0\rangle
    + \langle A_0B_1\rangle
    + \langle A_1B_0\rangle
    - \langle A_1B_1\rangle ,
\end{equation}
where $\langle A_{s}B_{t}\rangle$ denotes the expectation value of the joint observable determined by inputs $s$ and $t$. For uniformly sampled inputs $(s,t)$, the success probability is directly related to this correlator. For each input pair, the winning condition requires $\alpha\beta = (-1)^{st}$, and therefore
\begin{equation}
    P_{\mathrm{win}}(s,t)
    =
    \frac{1 + (-1)^{st}\langle A_{s}B_{t}\rangle}{2}.
\end{equation}
Averaging over the four input pairs gives
\begin{equation}
    P_{\mathrm{win}}
    =
    \frac{1}{4}\sum_{s,t\in\{0,1\}}
    \frac{1 + (-1)^{st}\langle A_sB_t\rangle}{2} =
    \frac{1}{2} + \frac{S}{8}.
\end{equation}
Thus, the classical strategy satisfies $S\leq 2$, corresponding to $P_{\mathrm{win}}\leq 3/4$, while the optimal quantum strategy reaches Tsirelson's bound $S=2\sqrt{2}$, corresponding to $P_{\mathrm{win}} = 1/2 + \sqrt{2}/4 = \cos^2(\pi/8)\simeq 0.853$.

The quantum bound is achieved by allowing Alice and Bob to share a maximally entangled two-qubit state and perform local measurements determined by $s_1,t_1\in\{0,1\}$. In observable form, Alice measures $A_0=Z$ or $A_1=X$, while Bob measures $B_0=(X+Z)/\sqrt{2}$ or $B_1=(Z-X)/\sqrt{2}$, where the subscripts denote the corresponding inputs $s_1$ and $t_1$, respectively. The measurement axes associated with $A_0$ and $A_1$ are separated by $\pi/2$ on the Bloch sphere, and the same holds for $B_0$ and $B_1$. This construction matches the optimal embedding  $\mathbf{x}^{o}$ used in the main text, in which $s_1$ and $t_1$ are locally encoded into the qubits of a pre-shared Bell pair. The $\mathbf{x}^{o}$ embedding can therefore be interpreted as a direct implementation of the optimal CHSH strategy, thereby enabling the DQML circuit to realize Tsirelson's bound. 

As shown in Table~\ref{tab:extended_tsirelson}, Bell-$1$ achieves an accuracy close to the Tsirelson bound, $P_{\mathrm{win}}\simeq 0.853$, whereas Bell-$0$ approaches the classical bound of $0.75$ under the optimal embedding. This close agreement with the known bounds validates our DQML circuit and indicates that the training procedure successfully recovers the optimal strategy.\\

\begin{figure}[b]
\centering
\includegraphics[width=0.45\linewidth]{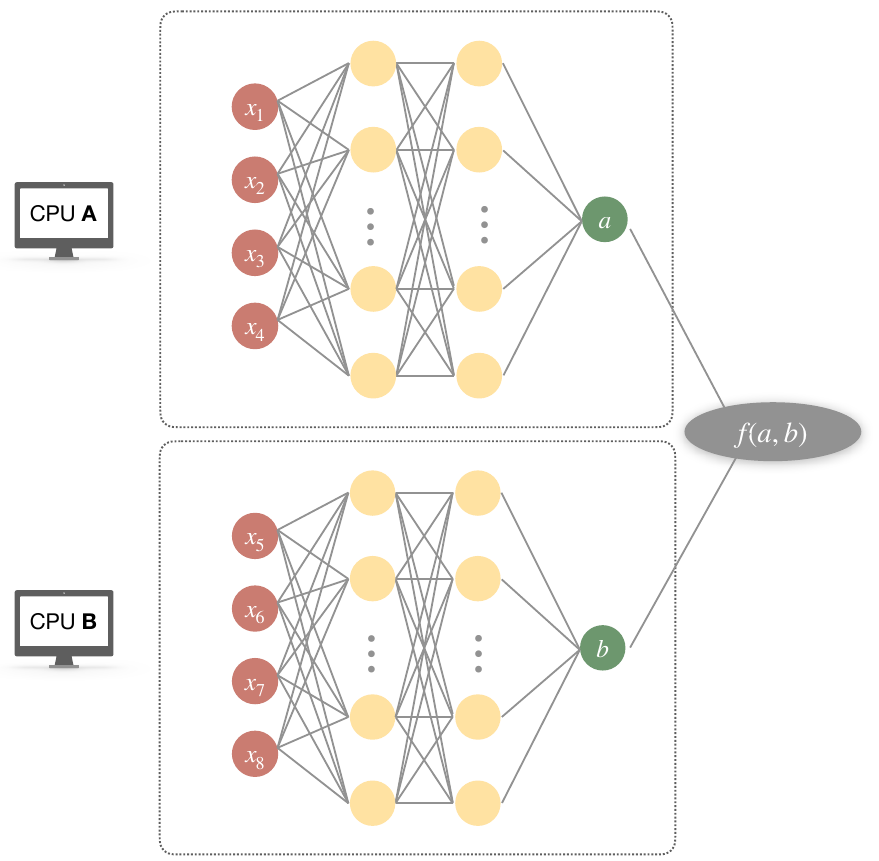}
\caption{\textbf{DNN schematic.}
The input data are split into two four-dimensional subsets, which are processed independently by CPU A and CPU B. Their outputs, $a$ and $b$, are combined in a final layer to produce the prediction $f(a,b)$. The model contains 246 trainable parameters, closely matching the depth-20 distributed QCNN benchmark with 242 parameters.}
\label{dnn}
\end{figure}

\begin{table}[t]
\centering
\begin{tabular}{c|ccc}
\hline
Dataset & DNN & Bell-0 & Bell-1 \\
\hline
A & $0.733 \pm 0.013$ & $0.739 \pm 0.007$ & $0.936 \pm 0.015$ \\
B & $0.782 \pm 0.008$ & $0.767 \pm 0.010$ & $0.916 \pm 0.020$ \\
C & $0.816 \pm 0.007$ & $0.804 \pm 0.008$ & $0.927 \pm 0.012$ \\
D & $0.850 \pm 0.007$ & $0.836 \pm 0.008$ & $0.941 \pm 0.016$ \\
E & $0.905 \pm 0.010$ & $0.895 \pm 0.008$ & $0.938 \pm 0.025$ \\
\hline
\end{tabular}
\caption{\textbf{Classification accuracy across five synthetic datasets.}
Standard deviations are obtained from 10 independent training runs. These values correspond to the result shown in Fig.~4\textbf{b} of the main text.}
\label{Belldataset}
\end{table}

\begin{table}[t]
\centering
\begin{tabular}{c|ccccc}
\hline
Depth & Bell-0 & Bell-1 & Bell-2 & Bell-3 & Bell-4 \\
\hline
5  & $0.721 \pm 0.008$ & $0.794 \pm 0.030$ & $0.851 \pm 0.011$ & $0.839 \pm 0.010$ & $0.778 \pm 0.010$ \\
10 & $0.730 \pm 0.004$ & $0.901 \pm 0.013$ & $0.917 \pm 0.010$ & $0.905 \pm 0.013$ & $0.806 \pm 0.005$ \\
15 & $0.752 \pm 0.022$ & $0.925 \pm 0.014$ & $0.946 \pm 0.010$ & $0.924 \pm 0.007$ & $0.812 \pm 0.004$ \\
20 & $0.749 \pm 0.025$ & $0.944 \pm 0.011$ & $0.961 \pm 0.005$ & $0.935 \pm 0.008$ & $0.814 \pm 0.008$ \\
25 & $0.746 \pm 0.004$ & $0.946 \pm 0.013$ & $0.965 \pm 0.006$ & $0.938 \pm 0.004$ & $0.813 \pm 0.003$ \\
\hline
\end{tabular}
\caption{\textbf{Classification accuracy for dataset A.} Standard deviations are obtained from 5 independent training runs. These values correspond to the result shown in Fig.~4\textbf{d} of the main text.}
\label{Belldepth}
\end{table}

\section{Supplementary Note 4 - Classification performance on synthetic datasets}
As a classical benchmark for the quantum convolutional neural network (QCNN) architecture~\cite{cong2019qcnn}, we introduce a distributed neural network (DNN). Because our synthetic datasets are low-dimensional, with eight input features, we employ a fully connected neural network rather than a convolutional neural network (CNN). As illustrated in Fig.~\ref{dnn}, the input layer consists of eight nodes. The upper four nodes correspond to the input vector $\mathbf{s}=(x_1,x_2,x_3,x_4)$ processed by CPU A, while the lower four nodes correspond to the input vector $\mathbf{t}=(x_5,x_6,x_7,x_8)$ processed by CPU B. No connections are introduced between the two branches before their final outputs are produced. 

The model consists of two hidden layers, with the number of nodes in each hidden layer chosen to be comparable to the number of input nodes, following established guidelines for network sizing~\cite{sheela2013review}. The total number of trainable parameters is 246, comparable to that of the depth-20 distributed QCNN with 242 parameters. After training, each branch is compressed into a single output, denoted by $a$ and $b$, analogous to the pooling operation in the distributed QCNN. To incorporate nonlinear interactions between the two branches, the final layer takes $a$, $b$, and their product $ab$ as inputs and combines them into the overall prediction $f(a,b)$. The network is trained using the same mean-squared-error (MSE) loss as in the DQML setting. 

Table~\ref{Belldataset} reports the validation accuracies of the DNN and DQML models in the Bell-$0$ and Bell-$1$ scenarios across five different datasets, corresponding to Fig.~4\textbf{b} of the main text. Each dataset is split into training and validation sets in a 3:1 ratio, and the accuracy is evaluated on the validation set according to Eq.~\eqref{supp:DQMLacc}. The reported values are obtained after 2,000 training iterations at convolutional-layer depth 20 and averaged over 10 independent runs. Across all datasets, the DNN shows performance comparable to the Bell-$0$ setting. Table~\ref{Belldepth} reports the validation accuracies of dataset A obtained in the Bell-$n$ scenarios, with $n=0,1,2,3,4$, for five different circuit depths, corresponding to Fig.~4\textbf{d} of the main text. All other settings are the same as those used for Table~\ref{Belldataset}, except that the reported values are averaged over 5 independent runs.\\

\begin{table}[t]
\centering
\begin{tabular}{c|cc}
\hline
Bell pairs & Mixing  & No mixing \\
\hline
0 & 0.738 $\pm$ 0.010 & 0.730 $\pm$ 0.004 \\
1 & 0.899 $\pm$ 0.036 & 0.901 $\pm$ 0.013 \\
2 & 0.922 $\pm$ 0.013 & 0.917 $\pm$ 0.010 \\
3 & 0.907 $\pm$ 0.011 & 0.905 $\pm$ 0.013 \\
4 & 0.903 $\pm$ 0.013 & 0.806 $\pm$ 0.005 \\
\hline

\end{tabular}
\caption{\textbf{Classification accuracy for dataset A with and without mixing layers.}
Standard deviations are obtained from 5 independent training runs for two configurations: three mixing layers followed by seven convolutional layers, and ten convolutional layers without mixing layers, corresponding to Fig.~5\textbf{e} of the main text.}

\label{tab:entanglement_mixing}
\end{table}

\begin{figure}[b]
\centering
\includegraphics[width=0.5\linewidth]{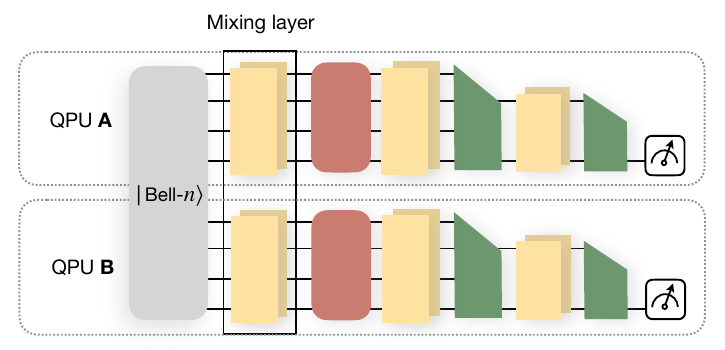}

\caption{\textbf{Entanglement-structure optimization.}
Entanglement mixing layers are inserted before the data-embedding circuit to optimize the structure of pre-shared entanglement. The overall circuit layout follows that of Fig.~2\textbf{c} in the main text.}  
\label{mixing}
\end{figure}

\section{Supplementary Note 5 - Entanglement restructuring}
To optimize the entanglement structure of pre-shared Bell pairs, as illustrated in Fig.~\ref{mixing}, we introduce additional trainable layers prior to the data-embedding circuit, referred to as entanglement mixing layers. These layers share the same circuit structure as the convolutional layers described in Supplementary Note~1. We compare two configurations: one with three mixing layers and seven 
convolutional layers, and one with ten convolutional layers and no mixing layers. The two configurations differ by 12 trainable parameters, yielding comparable circuit scales. The small difference arises from the convolutional layers applied after pooling, whereas mixing layers are introduced only before data embedding.

Table~\ref{tab:entanglement_mixing} shows the results corresponding to Fig.~4\textbf{e} of the main text, averaged over five independent trials. Within the standard deviation, Bell-1, Bell-2, and Bell-3 show negligible differences between the two configurations. By contrast, Bell-4 exhibits an improvement of approximately 10\% when mixing layers are included, bringing its performance in line with the other entangled configurations.

\begin{figure}[h]
\centering
\includegraphics[width=0.4\linewidth]{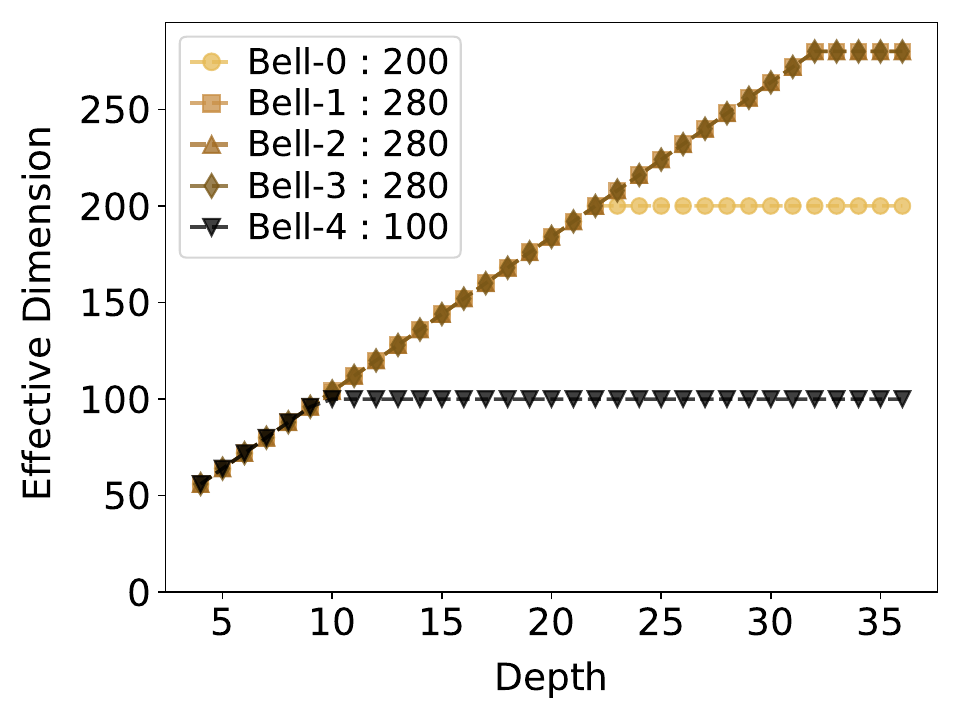}

\caption{\textbf{Effective dimension with entanglement mixing layers.}
Effective dimension as a function of total circuit depth for different Bell-$n$ settings, where 3 entanglement mixing layers and $d-3$ convolutional layers are included for depth $d$.}
\label{mix_eff}
\end{figure}

To further investigate this gain, we evaluate the effective dimension of the circuit with mixing layers following the procedure described in Supplementary Note~1, with the number of mixing layers fixed to three and the convolutional-layer depth varied. Fig.~\ref{mix_eff} shows the effective dimension as a function of total circuit depth, defined as the sum of mixing and convolutional layers, and Table~\ref{tab:effective_dim} lists the corresponding values alongside those without mixing layers. For Bell-1, 2, and 3, the inclusion of mixing layers increases the maximum achievable effective dimension. For Bell-4, however, the effective dimension converges to the same value as the no-mixing case once the depth exceeds 10. These results indicate that the performance gain of Bell-4 cannot be attributed to an increase in effective dimension, but rather reflects the role of entanglement structure in shaping the model's inductive bias.

\begin{table}[h]
\centering
\small
\begin{tabular}{c|ccccccccccccccc}
\hline
\multicolumn{16}{c}{No entanglement mixing layer} \\
\hline
Depth &1&2&3&4&5&6&7&8&9&10&11&12&13&14&15\\
\hline
Bell-0 &32&40&48&56&64&72&80&88&96&104&112&120&128&136&144\\
Bell-1 &32&40&48&56&64&72&80&88&96&104&112&120&128&136&144\\
Bell-2 &32&40&48&56&64&72&80&88&96&104&112&120&128&136&144\\
Bell-3 &32&40&48&56&64&72&80&88&96&104&112&120&128&136&144\\
Bell-4 &32&40&48&56&64&72&80&88&96&100&100&100&100&100&100\\
\hline
Depth &16&17&18&19&20&21&22&23&24&25&26&27&28&29&30\\
\hline
Bell-0 &152&160&168&176&184&192&200&200&200&200&200&200&200&200&200\\
Bell-1 &152&160&168&176&184&192&200&208&216&224&232&240&248&256&256\\
Bell-2 &152&160&168&176&184&192&200&208&216&224&232&240&248&256&256\\
Bell-3 &152&160&168&176&184&192&200&208&216&224&232&240&248&256&256\\
Bell-4 &100&100&100&100&100&100&100&100&100&100&100&100&100&100&100\\
\hline
\multicolumn{16}{c}{Entanglement mixing layer} \\
\hline
Depth &4&5&6&7&8&9&10&11&12&13&14&15&16&17&18\\
\hline
Bell-0 &56&64&72&80&88&96&104&112&120&128&136&144&152&160&168\\
Bell-1 &56&64&72&80&88&96&104&112&120&128&136&144&152&160&168\\
Bell-2 &56&64&72&80&88&96&104&112&120&128&136&144&152&160&168\\
Bell-3 &56&64&72&80&88&96&104&112&120&128&136&144&152&160&168\\
Bell-4 &56&64&72&80&88&96&100&100&100&100&100&100&100&100&100\\
\hline
Depth &19&20&21&22&23&24&25&26&27&28&29&30&31&32&33\\
\hline
Bell-0 &176&184&192&200&200&200&200&200&200&200&200&200&200&200&200\\
Bell-1 &176&184&192&200&208&216&224&232&240&248&256&264&272&280&280\\
Bell-2 &176&184&192&200&208&216&224&232&240&248&256&264&272&280&280\\
Bell-3 &176&184&192&200&208&216&224&232&240&248&256&264&272&280&280\\
Bell-4 &100&100&100&100&100&100&100&100&100&100&100&100&100&100&100\\
\hline
\end{tabular}
\caption{\textbf{Comparison of effective dimension.}
Effective dimensions for different Bell-$n$ settings as a function of circuit depth, shown for circuits without mixing layers and for circuits with three fixed entanglement mixing layers.}
\label{tab:effective_dim}
\end{table}




%% file: Bib.bib
@article{bennett1996purification,
   title={Purification of noisy entanglement and faithful teleportation via noisy channels},
  author={Bennett, Charles H and Brassard, Gilles and Popescu, Sandu and Schumacher, Benjamin and Smolin, John A and Wootters, William K},
  journal={Phys. Rev. Lett.},
  volume={76},
  number={5},
  pages={722},
  year={1996},
  publisher={APS}
}

@article{yin2017satellite,
  title={Satellite-based entanglement distribution over 1200 kilometers},
  author={Yin, Juan and Cao, Yuan and Li, Yu-Huai and Liao, Sheng-Kai and Zhang, Liang and Ren, Ji-Gang and Cai, Wen-Qi and Liu, Wei-Yue and Li, Bo and Dai, Hui and others},
  journal={Science},
  volume={356},
  number={6343},
  pages={1140--1144},
  year={2017},
  publisher={American Association for the Advancement of Science}
}

@article{wengerowsky2020passively,
  title={Passively stable distribution of polarisation entanglement over 192 km of deployed optical fibre},
  author={Wengerowsky, S{\"o}ren and Joshi, Siddarth Koduru and Steinlechner, Fabian and Zichi, Julien R and Liu, Bo and Scheidl, Thomas and Dobrovolskiy, Sergiy M and Molen, Ren{\'e} van der and Los, Johannes WN and Zwiller, Val and others},
  journal={npj Quantum Inf.},
  volume={6},
  number={1},
  pages={5},
  year={2020},
  publisher={Nature Publishing Group UK London}
}

@article{brassard2003quantum,
  title={Quantum communication complexity},
  author={Brassard, Gilles},
  journal={Found. Phys.},
  volume={33},
  number={11},
  pages={1593--1616},
  year={2003},
  publisher={Springer}
}

@article{abbas2021qnn,
  title={The power of quantum neural networks},
  author={Abbas, Amira and Sutter, David and Zoufal, Christa and Lucchi, Aur{\'e}lien and Figalli, Alessio and Woerner, Stefan},
  journal={Nat. Comput. Sci.},
  volume={1},
  number={6},
  pages={403--409},
  year={2021},
  publisher={Nature Publishing Group US New York}
}

@article{kimble2008quantuminternet,
  title={The quantum internet},
  author={Kimble, H Jeff},
  journal={Nature},
  volume={453},
  number={7198},
  pages={1023--1030},
  year={2008},
  publisher={Nature Publishing Group}
}

@article{cong2019qcnn,
  title={Quantum convolutional neural networks},
  author={Cong, Iris and Choi, Soonwon and Lukin, Mikhail D},
  journal={Nat. Phys.},
  volume={15},
  number={12},
  pages={1273--1278},
  year={2019},
  publisher={Nature Publishing Group UK London}
}

@article{bergholm2018pennylane,
  title={Pennylane: {A}utomatic differentiation of hybrid quantum-classical computations},
  author={Bergholm, Ville and Izaac, Josh and Schuld, Maria and Gogolin, Christian and Ahmed, Shahnawaz and Ajith, Vishnu and Alam, M Sohaib and Alonso-Linaje, Guillermo and AkashNarayanan, Bharath and Asadi, Ali and others},
  journal={arXiv:1811.04968},
  year={2018}
 
}

@article{hur2022quantum,
  author  = {Hur, Tak and Kim, Leeseok and Park, Daniel K.},
  title   = {Quantum convolutional neural network for classical data classification},
  journal = {Quantum Mach. Intell.},
  volume  = {4},
  number  = {1},
  pages   = {3},
  year    = {2022},
  doi     = {10.1007/s42484-021-00061-x},
  
}

@article{havlivcek2019supervised,
  title={Supervised learning with quantum-enhanced feature spaces},
  author={Havl{\'\i}{\v{c}}ek, Vojt{\v{e}}ch and C{\'o}rcoles, Antonio D and Temme, Kristan and Harrow, Aram W and Kandala, Abhinav and Chow, Jerry M and Gambetta, Jay M},
  journal={Nature},
  volume={567},
  number={7747},
  pages={209--212},
  year={2019},
  publisher={Nature Publishing Group UK London}
}

@article{haug2021capacity,
  title={Capacity and Quantum Geometry of Parametrized Quantum Circuits},
  author={Haug, Tobias and Bharti, Kishor and Kim, MS},
  journal={PRX Quantum},
  volume={2},
  number={4},
  pages={040309},
  year={2021},
  publisher={APS}
}

@article{cirel1980quantum,
  title={{Quantum generalizations of Bell's inequality}},
  author={Cirel'son, Boris S},
  journal={Lett. Math. Phys.},
  volume={4},
  number={2},
  pages={93--100},
  year={1980},
  publisher={Springer}
}

@article{bowles2023contextuality,
  title={Contextuality and inductive bias in quantum machine learning},
  author={Bowles, Joseph and Wright, Victoria J and Farkas, M{\'a}t{\'e} and Killoran, Nathan and Schuld, Maria},
  journal={arXiv:2302.01365},
  year={2023}
}

@article{preskill2018NISQ,
  doi = {10.22331/q-2018-08-06-79},
  title = {Quantum computing in the {NISQ} era and beyond},
  author = {Preskill, John},
  journal = {{Quantum}},
  issn = {2521-327X},
  publisher = {{Verein zur F{\"{o}}rderung des Open Access Publizierens in den Quantenwissenschaften}},
  volume = {2},
  pages = {79},
  month = aug,
  year = {2018}
}

@article{wehner2018quantuminternet,
  title={Quantum internet: {A} vision for the road ahead},
  author={Wehner, Stephanie and Elkouss, David and Hanson, Ronald},
  journal={Science},
  volume={362},
  number={6412},
  pages={eaam9288},
  year={2018},
  publisher={American Association for the Advancement of Science}
}

@article{caleffi2024distributed,
title={Distributed quantum computing: A survey},
  author={Caleffi, Marcello and Amoretti, Michele and Ferrari, Davide and Illiano, Jessica and Manzalini, Antonio and Cacciapuoti, Angela Sara},
  journal={Comput. Netw.},
  volume={254},
  pages={110672},
  year={2024},
  publisher={Elsevier}
}

@article{barral2025review,
  title={Review of distributed quantum computing: From single {QPU} to high performance quantum computing},
  author={Barral, David and Cardama, F Javier and D{\'\i}az-Camacho, Guillermo and Fa{\'\i}lde, Daniel and Llovo, Iago F and Mussa-Juane, Mariamo and V{\'a}zquez-P{\'e}rez, Jorge and Villasuso, Juan and Pi{\~n}eiro, C{\'e}sar and Costas, Natalia and others},
  journal={Comput. Sci. Rev.},
  volume={57},
  pages={100747},
  year={2025},
  publisher={Elsevier}
}

@article{main2025distributed,
  title={Distributed quantum computing across an optical network link},
  author={Main, Dougal and Drmota, Peter and Nadlinger, David P and Ainley, Ellis M and Agrawal, Ayush and Nichol, Bethan C and Srinivas, Raghavendra and Araneda, Gabriel and Lucas, David M},
  journal={Nature},
  volume={638},
  number={8050},
  pages={383--388},
  year={2025},
  publisher={Nature Publishing Group UK London}
}

@article{carrera2024combining,
  title={Combining quantum processors with real-time classical communication},
  author={Carrera Vazquez, Almudena and Tornow, Caroline and Rist{\`e}, Diego and Woerner, Stefan and Takita, Maika and Egger, Daniel J},
  journal={Nature},
  volume={636},
  number={8041},
  pages={75--79},
  year={2024},
  publisher={Nature Publishing Group UK London}
}

@article{aghaee2025scaling,
  title={Scaling and networking a modular photonic quantum computer},
  author={Aghaee Rad, H and Ainsworth, Thomas and Alexander, Rafael N and Altieri, Brandon and Askarani, Mohsen F and Baby, R and Banchi, Leonardo and Baragiola, Ben Q and Bourassa, J Eli and Chadwick, RS and others},
  journal={Nature},
  volume={638},
  number={8052},
  pages={912--919},
  year={2025},
  publisher={Nature Publishing Group UK London}
}

@article{CHSH1969,
  title = {Proposed Experiment to Test Local Hidden-Variable Theories},
  author = {Clauser, John F. and Horne, Michael A. and Shimony, Abner and Holt, Richard A.},
  journal = {Phys. Rev. Lett.},
  volume = {23},
  issue = {15},
  pages = {880--884},
  numpages = {0},
  year = {1969},
  month = {Oct},
  publisher = {American Physical Society},
    doi = {https://doi.org/10.1103/PhysRevLett.23.880}
}

@article{knaut2024entanglement,
  title={Entanglement of nanophotonic quantum memory nodes in a telecom network},
  author={Knaut, Can M and Suleymanzade, Aziza and Wei, Y-C and Assumpcao, Daniel R and Stas, P-J and Huan, Yan Qi and Machielse, Bartholomeus and Knall, Erik N and Sutula, Madison and Baranes, Gefen and others},
  journal={Nature},
  volume={629},
  number={8012},
  pages={573--578},
  year={2024},
  publisher={Nature Publishing Group UK London}
}

@article{buhrman2010nonlocality,
  title = {Nonlocality and communication complexity},
  author = {Buhrman, Harry and Cleve, Richard and Massar, Serge and de Wolf, Ronald},
  journal = {Rev. Mod. Phys.},
  volume = {82},
  issue = {1},
  pages = {665--698},
  numpages = {0},
  year = {2010},
  month = {Mar},
  publisher = {American Physical Society},
  doi = {10.1103/RevModPhys.82.665}
}

@article{bennett1993teleporting,
  title={{Teleporting an unknown quantum state via dual classical and Einstein-Podolsky-Rosen channels}},
  author={Bennett, Charles H and Brassard, Gilles and Cr{\'e}peau, Claude and Jozsa, Richard and Peres, Asher and Wootters, William K},
  journal={Phys. Rev. Lett.},
  volume={70},
  number={13},
  pages={1895},
  year={1993},
  publisher={APS}
}

@article{zhou2025kilometer,
  title={A kilometer photonic link connecting superconducting circuits in two dilution refrigerators},
  author={Zhou, Yiyu and Wu, Yufeng and Li, Chunzhen and Shen, Mohan and Yang, Likai and Xie, Jiacheng and Tang, Hong X},
  journal={  arXiv:2508.02444},
  year={2025}
}

@article{ho2022entanglement,
  title={Entanglement-based quantum communication complexity beyond {B}ell nonlocality},
  author={Ho, Joseph and Moreno, George and Brito, Samura{\'\i} and Graffitti, Francesco and Morrison, Christopher L and Nery, Ranieri and Pickston, Alexander and Proietti, Massimiliano and Rabelo, Rafael and Fedrizzi, Alessandro and others},
  journal={npj Quantum Inf.},
  volume={8},
  number={1},
  pages={13},
  year={2022},
  publisher={Nature Publishing Group UK London}
}

@article{brunner2014bell,
title={Bell nonlocality},
  author={Brunner, Nicolas and Cavalcanti, Daniel and Pironio, Stefano and Scarani, Valerio and Wehner, Stephanie},
journal = {Rev. Mod. Phys.},
  volume={86},
  number={2},
  pages={419--478},
  year={2014},
  publisher={APS}
}

@article{buhrman2016quantum,
  title={Quantum communication complexity advantage implies violation of a {B}ell inequality},
  author={Buhrman, Harry and Czekaj, {\L}ukasz and Grudka, Andrzej and Horodecki, Micha{\l} and Horodecki, Pawe{\l} and Markiewicz, Marcin and Speelman, Florian and Strelchuk, Sergii},
  journal={Proc. Natl. Acad. Sci. U.S.A.},
  volume={113},
  number={12},
  pages={3191--3196},
  year={2016},
  publisher={National Academy of Sciences}
}

@article{brukner2004bells,
 title={Bell’s inequalities and quantum communication complexity},
  author={Brukner, {\v{C}}aslav and {\.Z}ukowski, Marek and Pan, Jian-Wei and Zeilinger, Anton},
  journal={Phys. Rev. Lett.},
  volume={92},
  number={12},
  pages={127901},
  year={2004},
  publisher={APS}
}

@article{hwang2025DQML,
  title={Distributed quantum machine learning via classical communication},
  author={Hwang, Kiwmann and Lim, Hyang-Tag and Kim, Yong-Su and Park, Daniel K and Kim, Yosep},
  journal={Quantum Sci. Technol.},
  volume={10},
  number={1},
  pages={015059},
  year={2025},
  publisher={IOP Publishing}
}

@article{saha2025highfidelity,
  title={High-fidelity remote entanglement of trapped atoms mediated by time-bin photons},
  author={Saha, Sagnik and Shalaev, Mikhail and O’Reilly, Jameson and Goetting, Isabella and Toh, George and Kalakuntla, Ashish and Yu, Yichao and Monroe, Christopher},
  journal={Nat. Commun.},
  volume={16},
  number={1},
  pages={2533},
  year={2025},
  publisher={Nature Publishing Group UK London}
}

@article{neumann2022continuous,
  title={Continuous entanglement distribution over a transnational 248 km fiber link},
  author={Neumann, Sebastian Philipp and Buchner, Alexander and Bulla, Lukas and Bohmann, Martin and Ursin, Rupert},
  journal={Nat. Commun.},
  volume={13},
  number={1},
  pages={6134},
  year={2022},
  publisher={Nature Publishing Group UK London}
}

@article{gao2022enhancing,
  title={Enhancing generative models via quantum correlations},
  author={Gao, Xun and Anschuetz, Eric R and Wang, Sheng-Tao and Cirac, J Ignacio and Lukin, Mikhail D},
  journal={Phys. Rev. X},
  volume={12},
  number={2},
  pages={021037},
  year={2022},
  publisher={APS}
}

@article{yam2025cryogenicMicrowaveLink,
  title={Cryogenic microwave link for quantum local area networks},
  author={Yam, W. K. and Renger, M and Gandorfer, S and Fesquet, F and Handschuh, M and Honasoge, KE and Kronowetter, F and Nojiri, Y and Partanen, M and Pfeiffer, M and others},
  journal={npj Quantum Inf.},
  volume={11},
  number={1},
  pages={87},
  year={2025},
  publisher={Nature Publishing Group UK London}
}

@article{stolk2024metropolitan,
  title={Metropolitan-scale heralded entanglement of solid-state qubits},
  author={Stolk, Arian J and van der Enden, Kian L and Slater, Marie-Christine and te Raa-Derckx, Ingmar and Botma, Pieter and Van Rantwijk, Joris and Biemond, JJ Benjamin and Hagen, Ronald AJ and Herfst, Rodolf W and Koek, Wouter D and others},
  journal={Sci. Adv.},
  volume={10},
  number={44},
  pages={eadp6442},
  year={2024},
  publisher={American Association for the Advancement of Science}
}

@article{lee2026distributed,
  title={Distributed photonic variational quantum eigensolver with parameterized weak measurements},
  author={Lee, Donghwa and Bilash, Bohdan and Lee, Jaehak and Lim, Hyang-Tag and Kim, Yosep and Lee, Seung-Woo and Kim, Yong-Su},
  journal={npj Quantum Inf.},
  volume={12},
  pages={20},
  year={2026},
  publisher={Nature Publishing Group UK London}
}

@article{gottesman1999demonstrating,
  title={Demonstrating the viability of universal quantum computation using teleportation and single-qubit operations},
  author={Gottesman, Daniel and Chuang, Isaac L},
  journal={Nature},
  volume={402},
  number={6760},
  pages={390--393},
  year={1999},
  publisher={Nature Publishing Group UK London}
}

@article{lei2023quantum,
  title={Quantum optical memory for entanglement distribution},
  author={Lei, Yisheng and Kimiaee Asadi, Faezeh and Zhong, Tian and Kuzmich, Alex and Simon, Christoph and Hosseini, Mahdi},
  journal={Optica},
  volume={10},
  number={11},
  pages={1511--1528},
  year={2023},
  publisher={Optica Publishing Group}
}

@article{wang2025nuclear,
  title={Nuclear spins in a solid exceeding 10-hour coherence times for ultra-long-term quantum storage},
  author={Wang, Fudong and Ren, Miaomiao and Sun, Weiye and Guo, Mucheng and Sellars, Matthew J and Ahlefeldt, Rose L and Bartholomew, John G and Yao, Juan and Liu, Shuping and Zhong, Manjin},
  journal={PRX Quantum},
  volume={6},
  number={1},
  pages={010302},
  year={2025},
  publisher={APS}
}

@article{azuma2023quantum,
title={Quantum repeaters: {F}rom quantum networks to the quantum internet},
  author={Azuma, Koji and Economou, Sophia E and Elkouss, David and Hilaire, Paul and Jiang, Liang and Lo, Hoi-Kwong and Tzitrin, Ilan},
  journal={Rev. Mod. Phys.},
  volume={95},
  number={4},
  pages={045006},
  year={2023},
  publisher={APS}
}

@article{wang2021single,
  title={Single ion qubit with estimated coherence time exceeding one hour},
  author={Wang, Pengfei and Luan, Chun-Yang and Qiao, Mu and Um, Mark and Zhang, Junhua and Wang, Ye and Yuan, Xiao and Gu, Mile and Zhang, Jingning and Kim, Kihwan},
  journal={Nat. Commun.},
  volume={12},
  number={1},
  pages={233},
  year={2021},
  publisher={Nature Publishing Group UK London}
}

@article{sheela2013review,
  author  = {Sheela, K. G. and Deepa, S. N.},
  title   = {Review on Methods to Fix Number of Hidden Neurons in Neural Networks},
  journal = {Math. Probl. Eng.},
  volume  = {2013},
  number  = {1},
  pages   = {425740},
  year    = {2013}
}

@article{anschuetz2023interpretable,
  title={Interpretable quantum advantage in neural sequence learning},
  author={Anschuetz, Eric R and Hu, Hong-Ye and Huang, Jin-Long and Gao, Xun},
  journal={PRX Quantum},
  volume={4},
  number={2},
  pages={020338},
  year={2023},
  publisher={APS}
}

@article{gross2009tooentangled,
  title={Most Quantum States Are Too Entangled To Be Useful As Computational Resources},
  author={Gross, David and Flammia, Steve T and Eisert, Jens},
  journal={Phys. Rev. Lett.},
  volume={102},
  number={19},
  pages={190501},
  year={2009},
  publisher={APS}
}

@article{bremner2009randomstates,
   title={Are Random Pure States Useful for Quantum computation?},
  author={Bremner, Michael J and Mora, Caterina and Winter, Andreas},
  journal={Phys. Rev. Lett.},
  volume={102},
  number={19},
  pages={190502},
  year={2009},
  publisher={APS}
}

@article{hayden2006generic,
  title={Aspects of generic entanglement},
  author={Hayden, Patrick and Leung, Debbie W and Winter, Andreas},
  journal={Commun. Math. Phys.},
  volume={265},
  number={1},
  pages={95--117},
  year={2006},
  publisher={Springer}
}

@article{piveteau2022entanglementsimple,
  author  = {Piveteau, Amélie and Pauwels, Jef and Håkansson, Emil and Muhammad, Sadiq and Bourennane, Mohamed and Tavakoli, Armin},
  title   = {Entanglement-assisted quantum communication with simple measurements},
  journal = {Nat. Commun.},
  volume  = {13},
  number  = {1},
  pages   = {7878},
  year    = {2022},
  doi     = {10.1038/s41467-022-33922-5},
  
}

@article{bharti2022noisy,
 title={Noisy intermediate-scale quantum algorithms},
  author={Bharti, Kishor and Cervera-Lierta, Alba and Kyaw, Thi Ha and Haug, Tobias and Alperin-Lea, Sumner and Anand, Abhinav and Degroote, Matthias and Heimonen, Hermanni and Kottmann, Jakob S and Menke, Tim and others},
  journal={Rev. Mod. Phys.},
  volume={94},
  number={1},
  pages={015004},
  year={2022},
  publisher={APS},
  doi = {https://doi.org/10.1103/RevModPhys.94.015004},
}
